\documentstyle[11pt,newpasp,twoside]{article}
\begin{document}
\setcounter{page}{3}
\addcontentsline{toc}{section}{From Historical Perspectives to Some
Modern Possibilities \\ \hspace{.25in} {\it L.H. Aller}}

\title{From Historical Perspectives to Some Modern Possibilities}
\author{Lawrence H. Aller}
\affil{Division of Astronomy \& Astrophysics, Dept.\ of Physics and
Astronomy, University of California, Los Angeles, CA 90095--1562} 

\begin{abstract}
A historical perspective on the study of asymmetries in planetary nebulae
(PNs) is presented. We also describe our ongoing work in high resolution
spectroscopy of planetaries, and discuss some likely future directions for
the study of asymmetrical PNs.
\end{abstract}

     The first systematic study of planetary nebulae (PNs) was undertaken at
Lick Observatory from 1914 to 1917 and is reported in detail in Volume 13 of
their publications. Of particular interest are the extensive papers of
Curtis (1918) on the shapes and forms of planetaries and of Campbell \&
Moore (1918) on their internal motions.

     Curtis' discussion was based on plates taken with the Crossley
Reflector with a plate scale of 38.7$''$ mm$^{-1}$.  Many PN are, of course,
quite compact so the fine details are lost because of the limited resolution
of the emulsion. Curtis obtained a series of graded exposures from which he
made drawings of each object, thus overcoming many of the limitations
imposed by the non-linearity of photographic emulsions and the limited
intensity range that can be accommodated by a single exposure. He recognized
that by going to a telescope of much longer focal length ``a wealth of
minute structural detail would be shown in features that appeared on
Crossley plates as diffuse areas and wisps.'' From direct photographs alone
(even without monochromatic images that showed widely different structures
in [O II], [O III], and He II, for example), Curtis noted: ``It is evident that
we have to do with structures of extraordinary complexities -- the aberrant
wisps and striae and other minor formal irregularities in such structures as
the Ring Nebula in Lyra, NGC 7009, NGC 7026 and others would seem
to defy all attempts to analyze the details, whatever hypothesis may be
adopted regarding the general form of the structure as a whole.''

     Curtis (1918) noted that the ring hypothesis fails, as we should expect
a large number of elliptical or edge-on forms. The hypothesis of ellipsoidal
shells of uniform thickness also fails, as it cannot explain very faint
central regions or faintness at the ends of the major axis. He proposed some
general classes: ellipsoids or sphere-ring forms, ring shells, ellipsoidal
shells, helical objects such as NGC 6543 and NGC 7293, and anomalous forms
like NGC 2440. He was unable to recognize such objects as NGC 7027 as shell
structures heavily obscured by dust and classified them as ``anomalous.''

     One of the earliest attempts to combine nebular direct photos with
kinematical data was that made by Warren K. Green (1917). He compared direct
images and radial velocity studies of NGC 6543 and NGC 7009 and tried to
interpret them with a theoretical picture of rotating shells of gas. He told
me that after he left Lick Observatory he wrote his thesis while he was in
the French Foreign Legion, and ``nevermore worked on planetary nebulae.''

     The idea that planetary nebulae had split and distorted spectral lines
because they were in rotation seriously impaired progress in this field for
many years. Offhand, starting from the level of understanding of these
objects at the beginning of the 20th century, the hypothesis has a taint of
plausibility. After all, from asteroids to galaxies, astronomical objects do
rotate, so why not planetary nebulae?  However expanding shells offer a
simpler, more rational explanation of the doubled lines, Perinne (1929)
suggested, while Zanstra (1931) soon conclusively demonstrated that only the
expansion hypothesis would work. The arguments are so elementary that there
is no need to review them here; basically, were the PNs in rotation, the
lines would be tilted when the slit was placed across the object
perpendicular to the rotation axis. Further, in NGC 7662, the He II
$\lambda$4686 and 5007 [O III] lines were mirror images of one another,
implying shells rotating in opposite directions. Nevertheless, Campbell and
Moore concluded that of the 23 PNs showing internal motions, 19 could be
interpretable as rotation!

     After this important initial effort, the next great leap forward came 
with the application of atomic physics to spectroscopy, and especially to 
the spectra of gaseous nebulae. The names of Zanstra, Bowen, Menzel, and 
Ambartzumian are particularly associated with this development, although 
many others played important supporting roles. By the forties we had firm 
ideas of how the spectra of gaseous nebulae could be interpreted and how 
measurement of the intensities of nebular lines could give important clues 
to their diagnostics, so important in understanding complex PNs.

The next important observational advances were made by Olin Wilson and
Rudolph Minkowski at the Mt. Wilson and Palomar Observatories. Wilson used
the coude spectrograph on the 100-inch telescope to observe the spectral
lines of many ions. With the aid of a multislit spectrograph, in which a
single slit was replaced of by a series of closely-spaced parallel slits, he
could observe the pattern of radial velocity motions over the entire
image. Thus the ``kinematical structure'' could be obtained across the whole
PN and in several spectral lines. Wilson found a common kinematical pattern
in many PNs in that the ions of highest excitation (e.g., [Ne V]) gave the
smallest expansion velocities, while those of low ionization (e.g.\ [O II])
gave the highest velocity of expansion (Wilson 1950). Possibly, radiation
pressure expelled the gases in the outermost part of the envelope outward
while pushing the inner part of the nebular shell backwards towards the
star. Not all PNs conform to this rule (see Sabbadin \& Hamazaoglu
1981). In some planetaries of low excitation, [O I] and [S II] show
substantial expansion velocities whilst other ions such as O$^+$, N$^+$ and
O$^{++}$ seem to show no expansion at all (e.g., IC 418; Wilson 1950).
Wilson also secured monochromatic images of IC 418 in H$\alpha$, [N II], and
[O III] from which the spatial distributions of the relevant ions could be
deduced and compared with theoretical predictions of stratification.

     Direct photographs of a number of PNs were secured at the Mt.\ Palomar
200$''$ telescope by Minkowski in the 1940s and 1950s. Unfortunately, the
published record of these observations is only fragmentary; see e.g.\
Minkowski (1964) for some selected images. Isophotic contours of a number of
images were also published by Aller (1956). The original plan by Minkowski
and Wilson to include both large scale images and kinematic data in what
might be considered an ``updated Lick Vol.\ 13'' was never implemented,
unfortunately.

     Minkowski obtained images in H$\alpha$ + [N II], 4340, [O II], [0 III] and
He II in various PNs. These images bear out the intricacies hinted at by
Curtis and show some remarkable differences between nebulae. Perhaps the
most dramatic comparison is between NGC 7293 (the faint Helix Nebula in
Aquarius), with its numerous famous ``cometary'' structures studied by many
observers, and the smooth NGC 3587 (Owl Nebula), which seemed to show no
fine structure at all.

     Special mention must be made of Minkowski \& Osterbrock's (1960)
observations of NGC 6720 and NGC 650-651, which appear to be cylinder or
ringlike forms seen in different projections on the sky (compare with Curtis
1918). They estimate the electron densities inside the ring to be lower than
in the rings from the [O II] $\lambda$3729/$\lambda$3726 line ratios. The
physical structures of the two PNs appear to be closely similar; the spatial
orientations differ.

     With the advent of the {\it Hubble Space Telescope} (HST) and adaptive
optics and the supplementing of direct images with high dispersion
spectroscopic data, considerable progress seems possible. We have used the
Hamilton Echelle Spectrograph at the coude focus of the Lick Observatory 3m
telescope to observe lines from 3660 to 10,125 \AA\ with a spectral
resolution generally of the order of 0.2\AA\ (full-width half maximum). The
slit length is generally taken as 4.0$''$ to avoid overlapping echelle
orders. These data often are supplemented by observations with the
International Ultraviolet Explorer (IUE).

     Note that the size of the slit generally used ($4.0'' \times 1.2''$) is
almost invariably smaller than the size of the nebula under investigation.
In praxi, this means that we can reach faint, closely packed lines, like the
[N I] $\lambda\lambda$5198, 5200 pair, at a particular point in the nebular
image but cannot explore the line variations from point to point. For this
purpose, long slit data such as those employed by, e.g., Sabbadin \&
Hamazaoglu (1982) must be
used. While our setup is suitable for dredging up faint features,
large-scale studies of spatial excitation variations require monochromatic
images or long-slit spectra.

In the 1940s it was recognized that the ratio of $\lambda$4363 to the
$\lambda\lambda$4959, 5007 [O III] lines could give a good clue to the
electron temperatures in gaseous nebulae (Menzel et al.\ 1941) and that the
$\lambda$3726/$\lambda$3729 [O II] line ratio would be valuable for getting
electron densities (Aller, Ufford, \& van Vleck 1949). What was needed were
good cross-sections for collisional excitations of the relevant metastable
levels. These were provided by Seaton (1954a,b) and by Seaton \& Osterbrock
(1957). Improvements in the quantum mechanical treatment have been made over
the years, so that a greater variety of information can be extracted from
the forbidden line data, particularly for [O II] and other ions with three
equivalent $p$ electrons.

By comparing lines of the nebular type transitions, e.g., $\lambda\lambda$6717
and 6730 [S II], with those of the transauroral type transitions,
$\lambda\lambda$4068, 4076 [S II] (or the auroral type transitions that in
this instance fall in the near-IR near 1 $\mu$m), one can obtain both the
electron density and temperature for the same strata. Recall that in the
earlier work on nebular diagnostics we obtained the electron temperature in
the [O III] (O$^{++}$) zone but the electron density in the [O II] (O$^{+}$)
zone. Now it is possible to obtain the temperature and density in the same
zone, e.g. O$^+$ (Keenan et al 1999). For the radiations of [N II], [O III],
[Ne V], involving two equivalent $p$ electrons, we need to compare optical
region nebular-type transitions with IR transitions. This step involves
comparing data in very different spectral regions secured with radically
different detectors.

     Calculation of collision cross sections involving the auroral and
transauroral jumps in systems with 3 equivalent $p$ electrons are difficult
and have been completed successfully only in recent years, but we can now
use lines of [O II], [Ne IV], [S II], and [Ar IV] to get diagnostics for nebular
regions of very different excitation levels (Keenan et al.\ 1996, 1997, 1998,
1999). With the improvement of measurements of infrared lines, especially
those measurements obtained above the earth's atmosphere, we can greatly
extend PN diagnostics. In spectra as well as images we are always dealing
with a two-dimensional projection of a three-dimensional image. The fine
structure of the nebula may be below the resolution of the imaging element
even when seeing is eliminated, as in HST data. And we are always taking some
kind of an average along the line of sight.

     That asymmetrical forms of PNs may be related to factors such as their 
chemical composition or mode of excitation is perhaps illustrated by 
objects such as NGC 6537. Its isophotic contours as measured on a plate 
secured by Minkowski are shown in Aller (1956; p.\ 243). This PN appears to 
show 4 arms with a strong increase in brightness towards the center. A 
possible interpretation by means of looped filaments is shown in the
cartoon accompanying the plate.

     The spectrum is most remarkable (see, e.g., Feibelman et al.\ 1985;
Aller et al.\ 1999). Rowlands et al.\ (1994) called attention to the
unusually high electron temperatures found in this object and in NGC 6302
($T_e\sim$ 41,000 K and $T_e >$60,000 K, respectively, as
deduced from [Ne V]). The electron temperatures range
from $\sim$6500 K for [S II] and $\sim$15,000 K for [O II] to $\sim$30,000 K
for [Ne IV] and, as we have mentioned, even higher for [Ne V]!  Many years
ago, Menzel \& Aller (1941) showed that in a photoionized nebula of
approximately solar composition, the electron temperature would not exceed
about 20,000 K, even though the central star temperature might be
$\ge$200,000 K. Clearly the high [Ne V] $T_e$ values found in NGC 6537 and
NGC 6302 must be due to some cause other than the photoionization mechanism
that controls most planetaries.

     Is it any surprise that PNs in which shock excitation plays an 
important role may be asymmetrical? Another example, whose spectrum we
are studying intensively, is NGC 6543 (the Cat's Eye Nebula).

     The importance of high resolution spatial resolution augmented by
intensive spectroscopic data is obvious. It is important that additional
data for the infra-red and ultraviolet be secured. The significance of
observations from above Earth's atmosphere is clear. But it is also
important that these efforts should be supported by concurrent theoretical
studies, involving crucial atomic parameters, as well as by
theoretical structural and hydrodynamical investigations.

\end{document}